Application of gold in the field of heterogeneous catalysis

Siwei Luo[1]
[1]Ohio State University, Columbus, OH 43210

**Abstract:**

Gold has been long thought as an inert metal which finds most of its use in jewelry and monetary exchange. However, catalysis by gold has rapidly become a hot topic in chemistry ever since Haruta and Hutchings found gold to be an extraordinary good heterogeneous catalyst in certain reactions. Here in this paper, several model reactions which made gold historically famous as a catalyst and a currently hot topic will be demonstrated, such as oxidation of CO, selective oxidation, and hydrodechlorination. Conclusions on the chemical nature of gold will be made as well as future perspectives of designing gold as a better catalyst.

**0. Introduction**

Gold has been a highly sought-after precious metal for coinage, jewelry, and other arts since the beginning of recorded history. Chemically, gold is a transition metal rich in coordination numbers. However, compared with other metals, bulk gold is chemically least reactive which hardly involves in any catalytic reactions. Even if any, gold was not showed to be better than its fellows, such as Pd, Pt, and Ag.[1-5] However, a breakthrough was triggered by Haruta and Hutchings independently and almost simultaneously in around 1987 that gold served as an extraordinary good heterogeneous catalyst in the low-temperature oxidation of CO and the hydrochlorination of ethyne to vinyl chloride, respectively. For the very first time, gold was showed to be the irreplaceable catalyst for these reactions, in sharp contrast to its poor catalytic performance before.

For both CO oxidation and hydrochlorination, the active gold constituent was down to nano-scale. Although gold in bulk is poorly active as a catalyst, it turns out to be uniquely active when deposited as nanoparticles (NPs) with diameters smaller than 5 nm on base metal oxides, carbons, and polymers. There are three important factors that define the catalysis by gold: selection of support materials, size and shape (contact structure) control. The characteristic features of gold catalysts are firstly ambient temperature activity, secondly promotion by water, and lastly unique selectivity which is usually different from that of Pd and Pt. Gold NPs deposited on semiconductor metal oxides such as $TiO_2$, $Co_3O_4$, and $Fe_2O_3$ exhibit markedly high catalytic activity for CO oxidation even at a temperature as low as 200K. They have already been commercially applied to an odor eater for rest rooms in Japan.

With the flourishing development of nanoscience, catalysis by gold has rapidly become a hot topic in chemistry, with a new discovery being made almost every week. Table 1 shows gold catalysts with selected materials can promote many reactions, which, when catalyzed by other metals, usually need much higher temperatures and lead to lower selectivity. Figure 1 shows the number of publications on gold catalysis from 1900 to 2005, indicating gold catalysis, especially heterogeneous catalysis has become a highly dynamic hot spot in the field of catalysis research.

Table 1. Reactions catalyzed by gold. Adapted from [1-5]

| Type of reaction | Reactant | Temperature (K) | Support | Notes |
| --- | --- | --- | --- | --- |
| Complete oxidation | CO | 200–400 | $Be(OH)_2$, $Mg(OH)_2$, $Mn_2O_3$, $Fe_2O_3$, etc. | acidic metal oxides are excluded as a support. |
|  | HCHO | 300–450 | $TiO_2$ | regenerable by sun light |
|  | $CH_3OH$ | 300–450 | $TiO_2$ | regenerable by sun light |
|  | $CH_4$, $C_3H_8$ | 450–650 | $Co_3O_4$ | as active as Pd, Pt catalysts |
|  | trimethylamine | 330–500 | $Fe_2O_3$, $NiFe_2O_4$ | commercialized for odor eater |
| Oxidative decomposition | chlorofluorocarbon | 550–823 | $Co_3O_4$, $Al_2O_3$, $LaF_3$ | $LaF_3$ for HCN Eynthesis |
|  | o-chlorophenol | 450–550 | $Fe_2O_3$ | integrated with $Pt/SnO_2 + Ir/La_2O_3$ |
|  | dioxin | 400–500 | $Fe_2O_3$ | integrated with $Pt/SnO_2 + Ir/La_2O_3$ |
| Reduction or decomposition of NOx | $NO + C_3H_6$ | 450–800 | $Al_2O_3$ | to $N_2$, mixed with $Mn_2O_3$ |
|  | $N_2O$ (+ $O_2$ + $H_2O$) | 500– | $Co_3O_4$ | to $N_2$ |
| Oxidation or Reduction of COx | $CO + H_2O$ | 400–500 | $TiO_2$, $ZrO_2$, $CeO_2$ | to $CO_2 + H_2$ |
|  | $CO + 2H_2$ | 400–500 | ZnO | to methanol |
|  | $CO_2 + 3H_2$ | 400–500 | ZnO | to methanol |
|  | $CO_2 + 3H_2$ | 400–500 | $TiO_2$ | to CO |
| Selective oxidation | $C_3H_6 + H_2 + O_2$ | 300–500 | $TiO_2$ (anatase), $Ti-SiO_2$ | to propylene oxide |
|  | $C_3H_8 + O_2 + H_2$ | 300–400 | $TiO_2$ (anatase) | to aceton |
|  | $C_4H_{10} + O_2 + H_2$ | 300–400 | $TiO_2$ (anatase) | to butanol |
|  | glycols | room temp. | activated carbon | to $\alpha$-hydroxy acids, liquid phase |
| Selective hydrogenation | CH≡CH | 400–500 | $Al_2O_3$ | to ethylene |
|  | $CH_2=CH-CH=CH_2$ | 400–500 | $Al_2O_3$, $SiO_2$, $TiO_2$ | to butenes |
|  | crotonaldehyde | 500–550 | ZnO | to crotyl alcohol |
|  | acrolein | 513–593 | $ZrO_2$ | to allylalcohol |
| Hydrochlorination | CH≡CH | 373–393 | $AuCl_3$/activated carbon | to vinyl chloride |

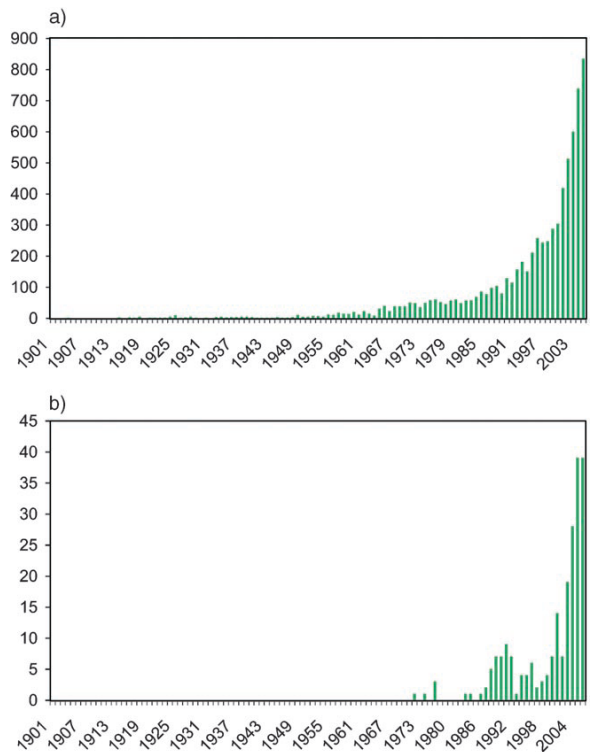

Figure 1. Publications on gold catalysis. a) Number of publications on "gold catalysis" from 1900 to May 2006. b) Number of publications on "homogeneous gold catalysis" only, showing how heterogeneous catalysis is still dominating. Adapted from [1-5]

## 1. Low temperature oxidation of CO

In the 1980s, Haruta et al. reported the supported gold catalysts were very active in the oxidation of CO even at temperatures significantly below 0°C (Figure 2). This surprisingly high activity, which has not been reached by other metals, is regarded as the very first potent evidence of gold's potential as a catalyst.

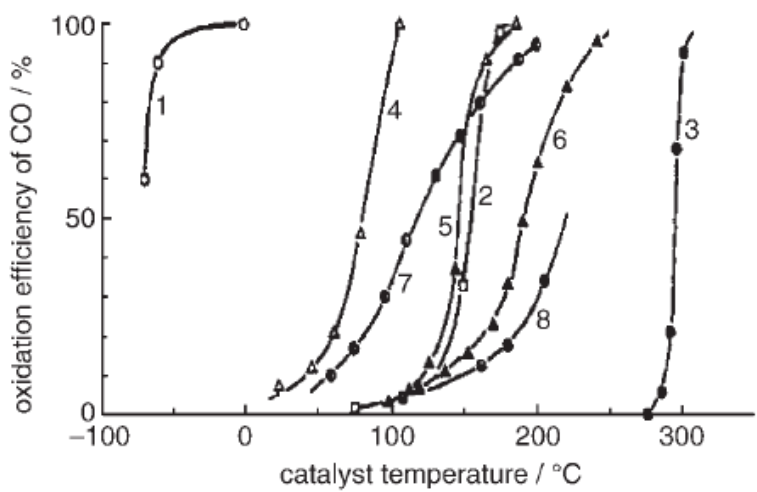

Figure 2. CO conversion over various catalysts as a function of temperature. Adapted from [1-5]

Certain principal factors have been agreed on for the high activity of supported Au NPs in low temperature oxidation of CO. However, the contribution of these factors is still disputing. Haruta and co-workers examined $TiO_2$ supported Pt and Au catalysts prepared with different methods, and showed that metal dispersion was dependent of preparation methods. According to the observation of Haruta, the deposition-precipitation method, which formed hemispherical metal particles adhering to the support (Figure 3), showed higher turnover frequency (TOF) than impregnation and photochemical deposition methods which formed spherical particles that have little interaction with the support. While for $Pt/TiO_2$, TOF was barely affected by the preparation methods, indicating only Pt involved in the reaction. Thus, the strong Au-support interaction must be responsible for the reaction mechanism of CO oxidation on $Au/TiO_2$.

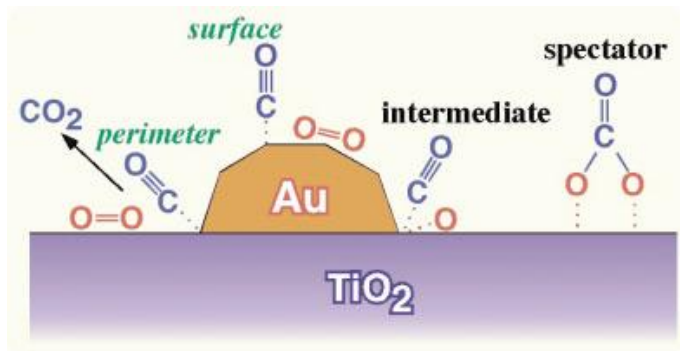

Figure 3. Schematic representation for CO oxidation pathways over Au/TiO2. Adapted from [1-5]

The impact of Au NPs size on the activity of CO oxidation was performed by both Haruta and Goodman and co-workers. Goodman discovered that Au/TiO$_2$ catalysts, either prepared by precipitation or by vapor-deposition, reach a maximum activity with the cluster size approximately at 3.0 nm (Figure 5). This is in great accordance with Haruta's founding of the size dependency on catalytic activity with Au on different supports (Figure 4).

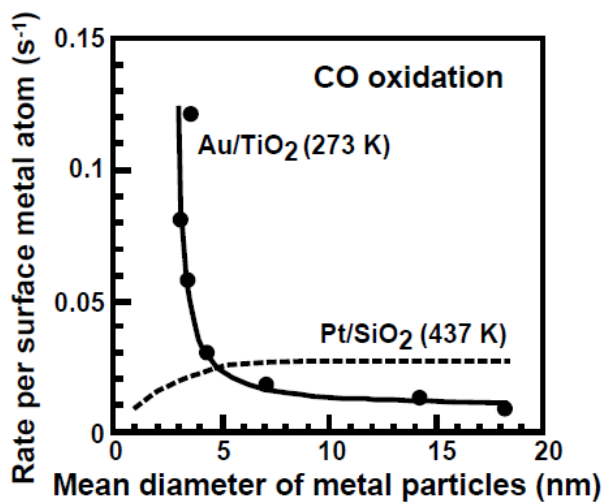

Figure 4. Turnover frequency (TOF) for CO oxidation over Au/TiO2 as a function of the mean diameter of Au particles. Adapted from [1-5]

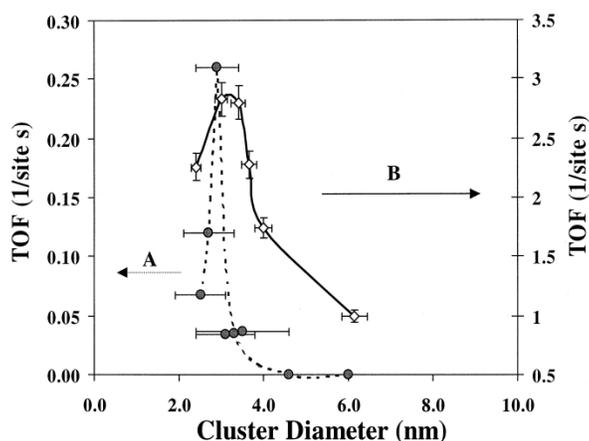

Fig 5. CO oxidation turnover frequencies _TOFs. as a function of the Au cluster size supported on TiO2 . Adapted from [1-5]

The nature and properties of oxide support, such as main group element oxides (Si, Mg, Sn), transition metal oxides (Fe, Ti, Ce) and mixed SiTiO gels, were examined by Grzyowska and co-workers. And transition metal oxide supports were found to be more active for CO oxidation than the main group element oxides.

Two different types of mechanisms for explaining the reaction were proposed. One involved only the Au NPs itself, while the other emphasized the interaction with the support. Although the dispute exists, a lot of studies have been done in kinetics, such as temperature studies and the kinetic isotope effect. Kinetic studies indicate CO and $O_2$ absorb competitively onto the surface and the surface reaction of CO and $O_2$ is the rate limiting step, with an experimentally determined rate expression of $\alpha[CO]^{0.05}[O_2]^{0.24}$. Furthermore, three temperature regions in the Arrhenius plot from 90K to 400K indicate three possible pathways for CO oxidation. Finally, water was proved to be irresponsible for the reaction mechanism due to the absence of a deuterium effects.

Although the convergence as to the mechanism for low temperature CO oxidation still exists due to the difficulty in quantifying variables such as preparation methods, types of supports, and the oxidation state of gold, the fact that supported gold catalysts are unique in the low temperature oxidation of CO will make gold very promising in the future industrial application like fuel cells in the presence of $H_2$, $H_2O$, and $CO_2$.[1-5]

## 2. Selective oxidation of alcohols and polyols

Alcohols and polyols provide a series of practical starting materials for producing a variety of chemicals as they can be obtained from natural and renewable sources in large amounts. In particular, selective oxidation can transform alcohols and polyols into the corresponding carbonylic or carboxylic derivatives which both generally represent attractive chemicals for organic synthesis.[1-5]

Selective oxidation of alcohols performed using oxygen in the presence of a catalytic system represents one of the most challenging reactions as it can be considered really attractive for its low environmental impact especially if compared to stoichiometric oxidation. Presently, there is a concurrence in competing properties among Ru, Pt, Pd and Au catalysts, gold based catalysts being among the most promising in terms of both selectivity and activity.[1-5]

Take selective oxidation of glycerol for example. Two research groups have been very active in oxidation of glycerol under mild conditions using gold based catalysts. Although the diversity of the possible reaction products (Figure 6) sets up a barrier of achieving high selectivity of our desired ones, Hutching's group proved the uniqueness of gold in selective oxidation. Using graphite as a support, in water solution at 60℃ and in the presence of NaOH, 100% selectivity to sodium glycerate could be readily achieved at 50-60% conversion.[1-5]

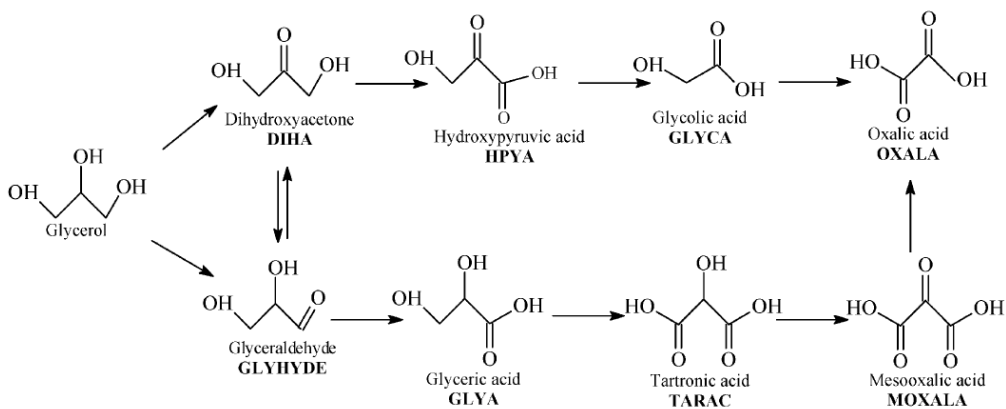

Figure 6. Reaction pathway for glycerol oxidation under basic conditions. Adapted from [1-5]

Another research group: Prati's group, carried out more detailed investigation of glycerol oxidation with gold. In his early study, the relationship between supported gold particles size and selectivity was established, indicating that the larger gold particle size (20nm), supported on suitable carbons, gave rise to low TOFs but very high selectivity of glycerate under mild conditions (30℃, 3 bar air). The yields were reported to be as high as 92%.

In his recent studies, another transition metal, such as Pt, and Pd, was incorporated into the gold catalysts as structured bimetallic NPs with the purpose of achieving both high activity and selectivity. The following foundings were reported:

1. Compared with monometallic catalysts (Pd, Pt, Au), bimetallic ones (Pd-Au, Pt-Au) increased the activity significantly, suggesting a synergistic effect between Au and the other two metals. Selectivity to desired products was found to be dependent of particle size, morphology of the bimetallic particles and support. High selectivity to glyceric acid was favored by gold, while palladium favored the further transformation of glyceric acid to tartronic acid, and platinum favored further more to glyconic acid. As the same with monometallic gold, larger particle size showed lower activity while higher selectivity than smaller particle size.

2. A volcano-shaped relationship between the activity or selectivity with the bimetallic atomic ratio in $Au_xPd_y/C$ catalysts was observed.

3. Activity and selectivity were affected by the nature of supports (carbon, graphite, $TiO_2$, $Ti/SiO_2$, $SiO_2$)

## 3. Gold in hydrodechlorination reactions

Groundwater contaminated by hazardous chlorinated compounds, especially chlorinated ethenes, continues to be a significant environmental problem in industrialized nations. Palladium-based materials have been shown to be very effective as hydrodechlorination catalysts for the removal of chlorinated ethenes and other related compounds. However, relatively low catalytic activity and a propensity for deactivation are significant issues that prevent their widespread use in groundwater remediation. A recent work [6-22] in Wong's lab shows that palladium-on-gold nanoparticles (Pd/Au NPs) catalyze the hydrodechlorination (HDC) of trichloroethene (TCE) in water, at room temperature, and in the presence of hydrogen, with the most active Pd/Au material found to be that was >10, >70, and >2000 times higher than monometallic Pd NPs, Pd/Al2O3, and Pd black, respectively.

## 4. Conclusion

Gold, when its size down to nano-scale, robust catalytic activity emerges, favoring a lot of reactions with both high conversion and selectivity, such as CO oxidation at low temperature, selective oxidation of propylene, alcohols and polyols, selective hydrogenation, hydrocholorination, hydrodechlorination, etc. The uniqueness of catalysis by gold, as concluded by Haruta, lies in: strong interaction between gold and the support, the nature of the support, the size of the gold particles. Furthermore, the incorporation of another metal, such as Pt or Pd, can have synergistic effect leading to both high conversion and selectivity. The uniqueness and synergy of gold make it a very important metal as a catalyst both in fundamental research where ultra small gold cluster is super active and in green chemistry where room temperature conversion of biomass and pollutants are crucial.